\documentstyle [12pt] {article}

\catcode`@=12
\begin{document}
\vspace{0.5in}
\oddsidemargin -.375in
\newcount\sectionnumber
\sectionnumber=0
\def\be{\begin{equation}}
\def\ee{\end{equation}}
\begin{flushright} UH-511-824-95\\April 1995\
\end{flushright}
\vspace {.5in}
\begin{center}
{\Large\bf Non-Leptonic two body decays of Charmed and ${\bf \Lambda_b} $
  Baryons  \\}
\vspace{.5in}
{\bf Alakabha Datta \\}
\vspace{.1in}
 {\it
Physics Department, University of Hawaii at Manoa, 2505 Correa
Road, Honolulu, HI 96822, USA.}\\
\vskip .5in
\end{center}

\vskip .1in
\begin{abstract}

We calculate the two body Cabibbo allowed non-leptonic  decays of charmed
 baryons $\Lambda_c$ and $\Xi_c$
  which involve  transitions of a heavy quark to a light quark
. We use data on the Cabbibo favoured non-leptonic decays
 $\Lambda_c \rightarrow \Lambda \pi^{+}$ and
 $\Lambda_c \rightarrow \Sigma^{+} \pi^{0 }$
 to obtain information on the form factors in the $c \rightarrow s $
transition. We also calculate the decay $\Lambda_c \rightarrow p \phi$.
 Using HQET the
  information on form factors from the $c \rightarrow s $ transition
is used to model the form factors in $b \rightarrow s $
 transition which are then used in the study of $
\Lambda_b \rightarrow J/ \psi \Lambda $
 decay.

\end{abstract}
\vskip .25in
\section{\bf Introduction}
There is now a fair amount of experimental
 data available on charmed baryon decays while
more data on bottom baryon decays will be available in the future and there
are already several calculations of these decays in the literature. A
crucial input in the calculation of the semi-leptonic as well as the
non-leptonic decays of charmed and bottom baryons are the hadronic form
factors. These form factors can be calculated is specific models like the
quark model or the MIT bag model \cite{NTC,tc}
. Another approach is to use HQET to find
relations among form factors for baryons containing a heavy quark. For
instance in the heavy-to-heavy transition of the type $\Lambda_b \rightarrow
 \Lambda_c $ all form factors are expressible in terms of one Isgur-Wise
function and a HQET mass parameter ${\bar \Lambda}$ up to order $1/m_Q $
 where $ m_Q $ is the  c or b quark mass. For a
heavy to light transition of the $\Lambda$ type baryon (light degrees in
spin 0 state), for example of the type $ \Lambda_c \rightarrow \Lambda $,
the
use of HQET in the limit $m_Q \rightarrow \infty $ allows one to express all
the form factors in terms of only two form factors \cite{mR}. Semileptonic
decay of $\Lambda_c $ has been studied in this limit \cite{Grin,KK} where
Ref.\cite{KK} in addition
 also assumes $1/m_s$ expansion for the semi-leptonic decay of
$\Lambda_c \rightarrow \Lambda $.

 In heavy to light transitions $1/m_Q$
corrections can be important, especially for the charm sector. Pure HQET
analysis of these $ 1/m_Q $ corrections in the heavy to light transitions does
not lead to  interesting phenomenology as there are too many
form factors and there is hardly any predictive power left \cite {Glin,Al}.
 However
in Ref. \cite{Al} it is shown  that using a combination of HQET and some
reasonable assumptions,
all the form factors up to $ 1/m_Q $ corrections can be expressed in terms of
only two form factors evaluated at maximum momentum transfer.
 A specific choice for the $q^2$ dependence of the form factors(e.g, a
monopole,dipole etc) can
be used
for the form factors to extrapolate to arbitrary values of the four momentum
transfer $ q^{2}$. In this model therefore there are two inputs, the zeroth
order form factors $F_1^{0}$ and $F_2^{0}$ at  maximum $q^{2}$ or
$\omega=v.v'=1$, where $v$ and $v'$ are the initial and final
 baryon velocities.
In this work we use a slightly
modified version of the  model for the form factors developed in
 Ref. \cite{Al} to
study the non-leptonic decays of charmed and bottom baryons.
To proceed with our calculations we need the zeroth
order form factors $F_1^{0}$ and $F_2^{0}$ at  maximum $q^{2}$ or
 alternately
$F_1^{0}$ and $r=F_2^{0}/F_1^{0}$ at  maximum $q^{2}$. The best place to
 fix these inputs would be from measurements of semi-leptonic decays.
For instance the asymmetry measurement in $\Lambda_c \rightarrow \Lambda l
\nu_{l} $ could be used to fix $r$. There are measurements of $\Lambda_c
\rightarrow \Lambda l^{+}
\nu_{l} $ form factors by the CLEO collaboration
 \cite {cleo} but the fit to data in these studies assumes
the KK model \cite {KK} for the form factors and hence is not general
enough for our use.

We next look into the data on non-leptonic decays of
charmed baryons. The theoretical description of these processes is model
dependent and to that extent an extraction of $F_1^{0}(\omega=1)$
 and $r=F_2^{0}(\omega=1)/F_1^{0}(\omega=1)$
 using non-leptonic data would also be model dependent. Using the current
algebra model we can use the value of the decay rate of $\Lambda_c
\rightarrow
\Sigma \pi^{0}$ to fix the non-factorizable contribution to the Cabibbo
favoured charmed baryon decays. Next, we can use
the values of the decay rate and asymmetry of $\Lambda_c \rightarrow
\Lambda \pi^{+} $ to fix $F_1^{0}$ and $r=F_2^{0}/F_1^{0}$ at
 $\omega=1$. We
 calculate the decay rates and asymmetries of the $\Lambda_c$ and
 the $\Xi_{c}$ charmed baryons
decaying
into an uncharmed baryon and a pseudoscalar or a vector meson. In our
calculations we use SU(3) symmetry to relate the form factors in the $c
\rightarrow s $
transition to $c \rightarrow u $ transitions.
 Using the flavour symmetry of HQET one can use the same inputs
$F_1^{0}(\omega=1)$ and $r$,
 extracted from the charm sector, in the bottom sector
to study the decays of the bottom baryon. Below we describe the basic
features of the current
algebra model that we employ in the calculation of the non-leptonic decays
of the charmed and bottom baryons.

The starting point of non-leptonic decay
 calculations is the QCD corrected weak
  Hamiltonian. This effective current$\times$current Hamiltonian gives rise
to the following quark diagrams \cite{CH} : the internal and
external W-emission diagrams, which
result in the factorizable contribution, and the W-exchange diagrams which
gives
rise to the non-factorizable contribution.
The W-annihilation diagram is absent in
baryon decay and the W-loop diagram
does not contribute to Cabibbo allowed decays.
 In the large $N_{c}$ limit the
non-factorizable contribution is no longer color suppressed because of
$N_{c}$ W-exchange diagrams . This combinatorial factor $N_{c}$ cancels a
similar factor in the denominator.

The factorizable part of the decay amplitude is expressed in terms of six
form factors. For the decay of the charmed baryon into an uncharmed baryon and
the light pseudoscalar, to a very good approximation, only two form factors
contribute for a pion in the final state
. When the pseudoscalar is replaced by a vector meson four of
these form factors contribute.
 We use the pole model to
 calculate the non-factorizable part. This model assumes that the
non-factorizable decay
amplitude receives contributions primarily  from one
particle intermediate states and these contributions then
 show up as simple poles in
the decay amplitude. The various intermediate single particle states are the
ground state positive parity baryons which contribute only to the parity
conserving amplitude, the parity violating amplitude being small\cite{gol}
.The parity violating amplitude may receive contribution from
excited negative parity baryons. In the limit that the momentum of the
pseudoscalar $q\rightarrow 0$, the parity violating piece of the
amplitude reduces to
the usual current commutator term of current algebra. Even though
in charmed baryon decay the final state pseudoscalar meson is not soft, we
will still work in the soft-meson limit and represent the parity
violating piece
of the amplitude by the current commutator term.
  It is important to note
that using SU(3) symmetry all the weak matrix element between the positive
parity
baryon states can be expressed in terms of only one matrix element and
therefore
in this model the non-factorizable contribution is completely determined by
one weak matrix element between positive parity ground state baryons.
 Hence
 the prediction for the asymmetry parameter for decays, which have no
factorizable contribution (eg, $\Lambda_{c}\rightarrow\Sigma^{+}\pi^{0}$),
 is independent of
the baryon-baryon weak matrix element and depends only on the baryon masses.

It is relevant to compare our model with some of the recent models employed
in the calculation of Cabibbo favoured charmed baryon decays. In our model
we use a completely different model for the form factors than has been used
in other models to calculate the factorizable piece of the decay amplitude.
Regarding the non-factorizable contributions, we have assumed that
the current commutator term  represents the parity violating
non-factorizable amplitude even
  in the case of
charmed baryon decays where the  pseudoscalar momentum $q$
 is far from zero.  Large
corrections to this current algebra results have been calculated in
Ref. \cite{TC} and
Ref. \cite{XuK}. However these corrections depend on the model used to
estimate the baryon to baryon weak matrix element and the corrections
calculated in Ref. \cite{TC} and Ref. \cite{XuK}
are quite different. Phenomenologically both
these calculations fail in their prediction of the asymmetry measured in
 the decay $\Lambda_{c}\rightarrow\Sigma^{0}\pi^{+}$. This is
also true for another recent calculation on non-leptonic charmed baryon decays
using a spectator quark model by K\"orner and Kramer \cite{KK1}. However
   the central
value of the measured asymmetry for the decay
$\Lambda_{c}\rightarrow\Sigma^{0}\pi^{+}$
 compares very well with the current algebra
prediction. This seems to indicate that, at least
for the decay, $\Lambda_{c}\rightarrow\Sigma^{0}\pi^{+}$,
the correction to the current algebra result is small.
 In the light of the experimental results we have therefore
adopted the position that the major contribution to the
 non-factorizable parity violating part of the amplitude
comes from the current algebra commutator term. The
advantage of
such a scenario is that the only parameter needed to specify
the non-factorizable
contribution is a single baryon-baryon matrix element which can
 be fixed from
the decay rate of a process like
$\Lambda_{c}\rightarrow\Sigma^{+}\pi^{0}$
 (which has no factorizable contribution) and we do not have to rely on
model dependent calculation of the weak matrix element.

The decay $\Lambda_c \rightarrow p \phi$ is Cabibbo suppressed and
has only factorizable contribution.  The same
form factors that characterize the $c \rightarrow u$ transition in Cabibbo
favoured decays can also be used for this decay.

For the $\Lambda_b$ decay we
ignore the non-factorizable contribution. For the form factors in this
decay we have used the same value of $F_1^{0}(\omega=1)$
and $F_2^{0}(\omega=1)$ used in charmed baryon decays as
$F_1^{0}(\omega=1)$
and $F_2^{0}(\omega=1)$ are the form factors for $m_Q \rightarrow \infty$
 at $\omega=1$ and
so by heavy flavour symmetry they are the same for the charm and
bottom sector.

The paper is organized in the following way. In the next section we outline
our model for the calculation of the various charmed and bottom baryon decays
while in the third section we present our results.

\section{\bf Model}

{\bf Non-Leptonic Decays:}
Here we develop the formalism for the Cabibbo favoured decay of a charmed
baryon into an uncharmed baryon and either a pseudoscalar or a vector
meson. This formalism will also be used in the decay
$\Lambda_b\rightarrow J/\psi \Lambda$. We
 start with the
decay of a charmed baryon into a baryon and a pseudoscalar. The amplitude
for such a decay can be written as
\be
M(B_{i}\rightarrow B_{f}P)= iu_{B_{f}}(A+B\gamma_{5})u_{B_{i}}\\
\ee
In the rest frame of the parent baryon the decay amplitude reduces to
\be
M(B_{i}\rightarrow B_{f}P)= i\chi_{B_{f}}(S+P {\bf \sigma.q})
\chi_{B_{i}}\\
\ee
where $\bf q$ is the unit vector along the direction of the daughter baryon
momentum and $S=\sqrt{(2m_{c}(E_{f}+m_{f})}A$ and
 $P=\sqrt{(2m_{c}(E_{f}-m_{f})}B$ with $E_{f}$ and $m_{f}$ referring to the
final baryon energy and mass. The decay rates and various asymmetries are
given by
\be
\Gamma=\frac{Q}{8\pi{m_{c}}^{2}}(|S|^2 +|P|^2);
\alpha=\frac{2Re(S^{*}P)}{|S|^2 +|P|^2} \quad;\quad
\beta=\frac{2Im(S^{*}P)}{|S|^2 +|P|^2}\quad and \quad
\gamma=\frac{|S|^2 -|P|^2}{|S|^2 +|P|^2}\\
\ee
where $Q$ is the magnitude of the three momentum of the decay products.
The starting point
of our dynamical analysis
is the QCD corrected effective weak Hamiltonian for Cabbibo favoured decays
\begin{equation}
H_{W}=\frac{G_F}{2\sqrt2}V_{cs}V_{ud}(c_{+}O_{+}+c_{-}O_{-}) \\
\end{equation}
with $ O_{\pm}=({\bar s}c)({\bar u}d)\pm({\bar s}d)({\bar u}c)$ where we have
omitted the Dirac structure $\gamma_{\mu}(1-\gamma_5)$ between the quark
fields inside each parentheses. $V_{cs}$ and $V_{ud}$ are the usual CKM
matrix elements while $c_{\pm}$ are the Wilson's coefficients evaluated
at the charm quark mass scale.
In our model we write the decay amplitude as
\be
M(B_{i}\rightarrow B_{f}P)=M(B_{i}\rightarrow B_{f}P)_{fac} +
M(B_{i}\rightarrow B_{f}P)_{nonfac}\\
\ee
{}From the structure of the Hamiltonian factorization occurs with a $\pi^{+}$
and ${\bar K}^{0}$ in the final state. The factorizable contribution is given
by
\begin{eqnarray}
M(B_{i}\rightarrow B_{f}\pi^{+}) &=&\frac{G_F}{2\sqrt2}V_{cs}V_{ud}
[c_1+\frac{c_2}{N_c}]<\pi^{+}|{\bar u}d|0><B_f|{\bar s}c|B_i>\\
M(B_{i}\rightarrow B_{f}{\bar K}^0)&=&\frac{G_F}{2\sqrt2}V_{cs}V_{ud}
[c_2+\frac{c_1}{N_c}]<{\bar K}^0|{\bar s}d|0><B_f|{\bar u}c|B_i>\
\end{eqnarray}
where $c_1=\frac{1}{2}(c_{+}+c_{-})$\quad;\quad $c_2=\frac{1}{2}(c_{+}-c_{-})$
with $N_c$ being the number of colors. The $N_c$ suppressed terms come from
the Fierz reordering of the operators $O_{\pm}$. For a satisfactory
descripotion of non-leptonic
decays of mesons it was found that the Fierz ordered contribution should be
 omitted
\cite{bauer}. This can be justified in the $1/N_c$ expansion method with
$N_c\rightarrow \infty$
\cite {buras}. We shall
therefore also work in the large $N_c$ limit. The matrix elements of the
current between baryonic states that appear in the equation above is
parametrized in terms of form factors. We define the six vector and axial
vector form factors through the following equations
\begin{eqnarray}
 \left < B'(p',s')
\mid \bar{q} \ \gamma^\mu \ Q \mid B_Q (p, s)
\right >
 =  \bar{u}_{B'} (p', s')
\left [f_1 \gamma^\mu - i \frac{f_2}{m_{B_c}}\sigma^{\mu\nu}
 q_\nu + \frac{f_3}{m_{B_c}} q^\mu \right ] u{_B{_Q}} (p, s)
\nonumber \\
 \left < B'(p', s')
\mid \bar{q} \ \gamma^\mu \gamma^5 \ Q \mid B_Q (p,s)
\right >
 =   \bar{u}_{B'} (p', s')
\left [g_1 \gamma^\mu - i \frac{g_2}{m_{B_c}}\sigma^{\mu\nu}
 q_\nu + \frac{g_3}{m_{B_c}} q^\mu \right ] \gamma^5 u{_B{_Q}} (p, s)
\end{eqnarray}
where $q^\mu = p^\mu-p'^\mu$ is the four momentum transfer, $B_Q$ is the
baryon with a heavy quark and $B'$ is the light baryon.
 In Ref. \cite{Al} we studied the form factors for  heavy
to light transitions involving baryons
 in HQET including corrections
up to $1/m_Q$ (Even though we studied charmed baryons in Ref. \cite{Al} the
results are applicable to the heavy to light transition of any $\Lambda$
type baryon containing a heavy quark).
 We found that
at $\omega=1$, in addition to the two zeroth order form factors
 form factors $F_1^{0}$ and $F_2^{0}$, there were five other unknown matrix
elements,four of which represent corrections from the chromomagnetic operator.
 In Ref.  \cite{Al} we made some assumptions about these unknowm
matrix elements and we were able to express all the form factors in terms
of two form factors $F_1^{0}$ and $F_2^{0}$. Without making any assumptions
about the corrections coming from the chromagnetic operator
 we write the form factors as
\begin{eqnarray}
\frac{f_1}{F^0_1} & = & 1+a + (m_{B_Q} + m_{B'})
    \left [ \frac{r+b/3}{2m_{B_Q}} - \frac{a'+b/3}{2m_{B'}} \right ]
\nonumber \\
\frac{f_2}{F^0_1 m_{B_Q}} & = & - \frac{r+b/3}{2m_{B_Q}} +
\frac{a''+b/3}{2m_{B'}}
\nonumber \\
\frac{f_3}{F^0_1 m_{B_Q}} &=& \frac{r+b/3}{2m_{B_Q}} + \frac{a'''+b/3}{2m_{B'}}
\nonumber \\
\frac{g_1}{F^0_1} &=& 1+r+\frac{2b}{3} - (m_{B_Q} - m_{B'}) \left [
\frac{r-a - \rho b/3}{2m_{B_{Q}}} + \frac{\rho b'/3}{2m_{B'}} \right ]
\nonumber \\
\frac{g_2}{F^0_1 m_{B_Q}} & = & - \frac{r-a-\rho b/3}{2m_{B_{Q}}} -
\frac{\rho b''/3}{2m_{B'}}
\nonumber \\
\frac{g_3}{F^0_1 m_{B_Q}} &=& \frac{r-a - \rho b/3}{2m_{B_Q}} -
\frac{\rho b'''/3}
{2m_{B'}}
\end{eqnarray}
where
\begin{eqnarray}
\bar{ \Lambda} &=& m_{B_Q}-m_Q \nonumber\\
\stackrel{\wedge}{m}_ {B'} &=& m_{B'}-m_q \nonumber\\
z_1 & = & (\bar{ \Lambda} + \stackrel{\wedge}{m}_ {B'}) \nonumber\\
z_2 & = & (\bar{ \Lambda} - \stackrel{\wedge}{m}_ {B'})\nonumber\\
 r & = & F^0_2/F^0_1 \nonumber\\
a & = & \frac{(z_1+\frac{4}{3}z_2)+r(z_1+\frac{1}{3}z_2)}{2m_Q} \nonumber\\
\rho & = & -\frac{6r}{1+r}{\frac{\stackrel{\wedge}{m}_ {B'}}{z_2}}
\nonumber\\
b & = & -\frac{(1+r)}{2m_Q}z_2 \
\end{eqnarray}
and $m_{B_Q}$ and $m_{B'}$ are the heavy and the light baryon masses while
$m_Q$ and $m_{q}$ are the masees of the heavy and the light quark
respectively.
The model
 used in this paper  corresponds to $a'=a''=0$,
$\rho b'=\rho b(1+2m_{B'}/m_{B_Q})$ and $\rho b''=\rho b$. The quantaties
$a'''$ and $\rho b'''$ are now calculable since the four matrix element
that represent the chromagnetic corrections are determined by our choice of
$a'$,$a''$,$\rho b'$ and $\rho b''$. The expressions
for $a'''$ and $\rho b'''$ are
\begin{eqnarray*}
a''' &=& a+a[1-\frac{2m_{B_Q}}{m_{B_Q}-m_{B'}} +\frac{\rho b}{a}
\frac{m_{B'}}{m_{B_Q}}]\nonumber\\
\frac{\rho'''b}{3} &= &\frac{\rho b}{3}+[(\frac{2m_{B'}}{3m_{B_Q}}
+\frac{4{m_{B'}}^{2}}{(m_{B_Q}-m_{B'})m_{B_Q}})\rho b
-\frac{4m_{B'}m_{B_Q}}{(m_{B_Q}-m_{B'})^{2}}a]\nonumber\\
\end{eqnarray*}
The
choice of the model described above is
dictated by the fact that it works well phenomenologically and the fact that
an expansion in $1/m_Q$ is valid. The condition for the
 validity of the $1/m_Q$ expansion is
defined through the
 constraint $|r|\leq 1$.
 To connect these assumptions with the
ones made in Ref. \cite{Al},
 we review the assumptions made about the corrections coming from the
chromomagnetic operator in Ref. \cite{Al}. We consider $m_{B'}/m_{B_Q}$
to be small and
 we relax some of the assumptions about
the chromomagnetic corrections in Ref. \cite{Al}.
  While we retain
$\delta F_1 + \delta F_2 + \delta F_3 = 0 $ (eqn.33 of Ref. \cite{Al}) we
only assume
(at $\omega=1$)
$\chi_{11}\sim\chi_{12}\sim \chi_{1}$ but do not constrain $\chi_{21}$
 and $\chi_{22}$.
 The above assumptions lead to $\chi_{1}(\omega=1)=x(\omega=1)/m_c $
\cite{Al}. The model for the form factors used in this work corresponds
to $\chi_{21} =\chi_{22}=\chi_{2}=-a$ in the limit $m_{B'}/m_{B_Q}$ is small.
 So we
see that the model employed here is almost identical to the model in
Ref. \cite{Al}(except for $\chi_{1}$ not equal to $\chi_{2}$)
in the limit $m_{B'}/m_{B_Q}$ is small.
  For the decays
of charmed baryons considered in this paper the difference between the two
models can be significant given the fact that $m_{B'}/m_{B_Q}$ is no longer
small. For bottom baryon decays we expect the two
models to yeild essentially identical results.
   Imposing
 the constraint on $r$ we find that we can fix $f$ and $g$ from the
measured asymmetry and decay rate of
$\Lambda_c \rightarrow \Lambda \pi ^{+}$. Taking into account
the experimental errors, the
  form factors f and g are such
that $(g-f)/g \leq 0.35$ if $ f<g $ and $(f-g)/f \leq 0.35$ if $ g<f $. Note
that in the $m_c \rightarrow \infty$ limit the form factors f
and g are equal. The inclusion of $1/m_c$ corrections destroys this
equality, and so the inequalities above represent the
size of the $1/m_c$ corrections. We also assume $F_1(\omega=1)^{0} >0$ in
our analysis.
The factorizable contributions to the decay amplitude can now be written as
\begin{eqnarray}
A_{fac}&=&\frac{G_F}{\sqrt2}V_{cs}V_{ud}f_{P}c_{k}[(m_f-m_i)f_{1}({m_{P}}^2)
+f_3({m_{P}}^2)\frac{{m_{P}}^2}{m_i}]
\nonumber\\
B_{fac} &= &\frac{G_F}{\sqrt2}V_{cs}V_{ud}f_{P}c_{k}[(m_f+m_i)g_{1}({m_{P}}^2)
+ g_3({m_{P}}^2)\frac{{m_{P}}^2}{m_i}]\
\end{eqnarray}
where $c_1(c_2)$ refer to $\pi^{+}$or ${\bar K}^{0}$ emission, $f_P$ is the
pseudoscalar decay constant and $f_1$ and $g_1$ are the form factors defined
in eqn.(8). In our analysis we shall use the SU(3) results
\begin{eqnarray}
{f_{1}}^{\Lambda_c \Lambda}&=&
\sqrt{\frac{2}{3}}{f_{1}}^{{\Xi_c}^{0A} {\Xi}^{0}}
\quad  \quad =-\sqrt{\frac{2}{3}}{f_{1}}^{{\Xi_c}^{+A} {\Xi}^{0}}
\quad  \quad =-\sqrt{\frac{2}{3}}{f_{1}}^{\Lambda_c p} \nonumber\\
&=& -\sqrt{\frac{2}{3}}{f_{1}}^{{\Xi_c}^{+A} {\Sigma}^{+}} \quad \quad
=\sqrt{\frac{4}{3}}{f_{1}}^{{\Xi_c}^{0A} {\Sigma}^{0}} \quad  \quad
=2{f_{1}}^{{\Xi_c}^{0A} {\Lambda}}\
\end{eqnarray}
It is important to note that strictly we should use the SU(3)
 relations for the zeroth
order form factors since the $1/m_Q$ corrections involve the baryon masses and
hence break SU(3), but this effect is small and is therefore neglected in
our analysis.

For the non-factorizable term we will use the pole model and current algebra
as outlined in the introduction. Following Ref. \cite{MRR} we write
the non-factorizable amplitude $R(k)$ as
\be
R(q)= R_{Born}(q)+{\bar R}(k)\
\ee
The usual approximation is to assume
\begin{eqnarray}
R(q,q^2=m_{P}^2)\simeq R_{Born}(q,q^2=m_{P}^2)+{\bar R}(0)\
\end{eqnarray}
Finally using reduction techniques for the amplitude one obtains
\begin{eqnarray}
R(q,q^2=m_{P}^2)\simeq \frac{-{\sqrt 2}}{f_P}<B|[Q_5,H^{PV}]|B_c> +
 R_{Born}(q,q^2=m_{P}^2)-{\bar R}(0)\
\end{eqnarray}
where $Q_{5}$ is the axial charge and $f_P$ is the pseudoscalar decay constant.

Clearly the first term in the amplitude above contributes to the parity
violating amplitude while the remaining terms
contribute to the parity conserving amplitude as the parity violating
amplitude is small \cite {gol}. (In the case of non-leptonic hyperon
decays $<B_f|H^{PV}|B_i>=0$ in the
SU(3) limit \cite {MRR}). Note in the case of charmed
baryon decays, as opposed to the hyperon decay case, ${\bar R}(0)$
is no longer small compared to $R_{Born}(q,q^2=m_{P}^2)$. Hence in our model
we have
\begin{eqnarray}
A &= &\frac{-{\sqrt 2}}{f_P}<B|[Q_5,H^{PV}]|B_c> \
\end{eqnarray}
\begin{eqnarray}
B=-[g_{B''B'P}\frac{<B''|H^{PC}|B>}{m_{B}-m_{B''}}\frac{m_{B}+m_{B'}}
{m_{B''}+m_{B'}}
+g_{BB'''P}\frac{<B'|H^{PC}|B'''>}{m_{B'}-m_{B'''}}
\frac{m_{B}+m_{B'}}
{m_{B}+m_{B'''}}]\
\end{eqnarray}
The first term in the expression for B is the s-channel
pole contribution while the
next term is the u-channel pole contribution.
 The strong pseudoscalar meson-baryon coupling $g_{B_i,B_j,P}$
can be related via the Goldberger-Treiman relation
 to the axial vector form
factors $ {g^A}_{B_i,B_j}$ as
\be
g_{B_i,B_j,P}=\frac{1}{f_P}(m_{B_i}+m_{B_j}){g^A}_{B_i,B_j}\
\ee
The axial form factors  $ {g^A}_{B_i,B_j}$ are of two types, those between
non-charmed baryons and those between charmed baryons. For the first type we
use  SU(3) parametrization with
\be
D+F=1.25 \quad;\quad D/F \approx 1.8 \
\ee
where the  D/F ratio is taken from a fit to hyperon semileptonic decay \cite
{DF}. The second type of form factors are between charmed baryons and
it is reasonable to use SU(4) symmetry and use the same D and F  is this
 case also. The justification for this lies in the fact the the transitions
are $\Delta C=0$ and so the baryon wavefunction mismatch in the overlap
integral is small\cite {XuK}. For the weak matrix element between the
positive parity baryons we will use the following SU(3) relation
\begin{eqnarray}
a_{{\Sigma}^{+}{\Lambda_c}^{+}}=a_{{\Sigma_c}^{0}{\Lambda}^{0}}
=a_{{\Xi}^{0A}{\Xi}^{0}}
=\sqrt{\frac{1}{3}}a_{{\Xi}^{0S}{\Xi}^{0}}
 =\sqrt{\frac{1}{3}}a_{{\Sigma_c}^{+}{\Sigma}^{+}}
=-\sqrt{\frac{1}{3}}a_{{\Sigma_c}^{0}{\Sigma}^{0}}\
\end{eqnarray}
where $ a_{B_i,B_f}=<B_f|H^{PC}|B_i>$. Using
 the above SU(3) relations the non-factorizable term is completely
specified in terms of one weak matrix element which we choose to be
$a_{{\Xi}^{0A}{\Xi}^{0}}$, and which we fix from the measured decay rate of
$\Lambda_c\rightarrow \Sigma^{+} \pi^{0}$.

For the decay where the meson in the final state is a vector meson we can
write the decay amplitude as
\begin{eqnarray}
M(B_{c}\rightarrow B_{f}V)=
iu_{B_{f}}\epsilon^{*\mu}[\gamma_{\mu}(a+b\gamma_{5})
+2(x+y\gamma_{5})P_{1\mu}]u_{B_{c}}\
\end{eqnarray}
where $P_{1\mu}$ is the four-momentum of the parent baryon and
$\epsilon^{*\mu}$ is the polarization of the vector meson. The kinematics
for this decay has been worked out in details in Ref. \cite{SPT}.
We can write down
the factorizable contribution as
\begin{eqnarray}
a_{fac} &=&\frac{G_F}{\sqrt2}V_{cs}V_{ud}f_{V}m_{V}c_{k}
[f_{1}({m_{V}}^2)+ \frac{m_f+m_i}{m_i}f_{2}({m_{V}}^2)]
\nonumber\\
b_{fac}&=&\frac{G_F}{\sqrt2}V_{cs}V_{ud}f_{V}m_{V}c_{k}
[g_{1}({m_{V}}^2)  + \frac{m_f-m_i}{m_i}g_{2}({m_{V}}^2)]\nonumber\\
x_{fac}&=&\frac{G_F}{\sqrt2}V_{cs}V_{ud}f_{V}m_{V}c_{k}
[\frac{f_{2}({m_{V}}^2)}{m_i}]\nonumber\\
y_{fac}&=&\frac{G_F}{\sqrt2}V_{cs}V_{ud}f_{V}m_{V}c_{k}
[\frac{g_{2}({m_{V}}^2)}{m_i}] \
\end{eqnarray}
where $c_1(c_2)$ refer to $\rho^{+}$or ${\bar K}^{*0}$ emmision, $f_V$ is the
vector meson decay constant, $m_{V}$
is the vector meson mass and $f_1,f_2$ and $g_1,g_2$ are the form
factors. For the pole term we will work in the approximation that $\rho$
generates isospin and so the couplings $g_{BBV}$ are pure F-type.
 Similar results apply to the
 decays $\Lambda_c\rightarrow p \phi$ and
 $\Lambda_b\rightarrow J/\psi\Lambda$ with the appropriate changes in
the QCD correction factor and the CKM matrix elements.

Before we present our results in the next section we list the various inputs
used in the calculations.
We begin with the calculations on the non-leptonic decays
of the charmed baryons.
As outlined in the introduction a fit to the decay rate and the asymmetry
for the decay $\Lambda_c \rightarrow \Lambda \pi^{+}$ is used to extract
$F_1^{0}(\omega=1)$ and $ r $.
The extracted values are $F_1^{0}(\omega=1)=0.46$ and $ r=-0.47 $. The values
for the Wilson's coefficients $c_1$ and $c_2$ were taken $\approx$
 1.32 and -0.59 respectively and
we have used $m_c=1.4$ GeV and $m_s=0.2$ GeV\cite{QM}.
 We found
that the non-factorizable contribution could be expressed in terms of the
single matrix element $a_{{\Xi}^{0A}{\Xi}^{0}}$. The measured decay rate of
$ \Lambda_c \rightarrow \Sigma^{+} \pi^{0} $
is used to extract $a_{{\Xi}^{0A}{\Xi}^{0}}=-5.48 \times 10^{-8} $
GeV. For
the vector meson decays we use, following Ref. \cite{TC},
$ f_{\rho}=f_{K^{*}}=0.221$ GeV. For the mode
$\Lambda_c \rightarrow p \phi$  we have used
$f_{\phi}=0.23$ GeV for the $\phi$ decay constant.
For the $\Lambda_b$
 decay we have
used $|V_{cb}|=0.040$ \cite{ne}, $c_2 \approx 0.23$, $f_{J/\psi}
=395$ MeV, and  pole masses
$m_V \cong 5.42$ GeV, $m_A \cong 5.86$ GeV\cite{NTC} . The
quark masses were taken as $m_b=
4.74$ GeV and $m_s=
0.20$ GeV \cite{QM}.

\section{\bf Results}
Starting with the results on the non-leptonic decays of the charmed baryons,
 in table. 1
 and 2 we give the predictions for the decay rates and asymmetry for the
non-leptonic decays $B_{i} \rightarrow B_f P $ and
$B_{i} \rightarrow B_f V $. In table. 3 we show the
 predictions for the mode $\Lambda_c \rightarrow p \phi$ and in table .4
we show the
predictions for
$ \Lambda_b \rightarrow J/ \psi \Lambda $.
 In table. 5 and table. 6 we show the form factors for
the $\Lambda_c\rightarrow \Lambda$ transition while in
table. 7 and table. 8 we show the form factors for
the $\Lambda_b\rightarrow \Lambda$ transition.

\begin{table}
\caption{Decay rates ($\times 10^{11}s^{-1}$), branching
ratios ($\times 10^{-3}$)
 and asymmetry predictions for Cabibbo favoured
$B_{i} \rightarrow B_f P $ decays. The asterisks indicate the input values.   }
\begin{center}
\begin{tabular}{|c|c|c|c|c|c|c|}
\hline
$ Process$ & $\Gamma_{Th}$ &$ BR_{Th}$ & $\Gamma_{Expt}$ & $
BR_{Expt}$ & $\alpha_{Th}$
& $\alpha_{Expt}$\\
\hline
$ \Lambda_c \rightarrow \Lambda \pi^{+} $ & $0.40$ & $7.9^{*}$ &
$ 0.40\pm 0.11$ & $ 7.9 \pm 0.18$
\cite{cleo 1} &
$ -0.94^{*} $ & $ -0.94^{+.22}_{-0.06} \cite{cleo 1}$ \\
\hline
$ \Lambda_c \rightarrow \Sigma^{0} \pi^{+} $ & $0.44 $ & $ 8.7^{*}$ &
$ 0.44 \pm 0.10$ & $ 8.7\pm 0.20$
\cite{PDG} &
$ -0.47 $ & $ -- $ \\
\hline
$ \Lambda_c \rightarrow \Sigma^{+} \pi^{0} $ & $0.44$ & $ 8.7$ &
 $ 0.44 \pm 0.12 \cite{cleo 1}$ & $ 8.7\pm 0.22 \cite{PDG}$ &
$ -0.47 $ & $ -0.45 \pm 0.31 \pm 0.06$\cite{cleo 1} \\
\hline
$ \Lambda_c \rightarrow p {\bar K}^{0} $ & $0.68$ & $ 13.4$ & $ 1.05\pm 0.20
$ & $ 21 \pm 0.4 \cite{PDG}$
 &
$ -0.91 $ & $ --$ \\
\hline
$ \Lambda_c \rightarrow \Xi^{0} { K}^{+} $ & $0.25$ & $4.9$ & $ 0.17\pm 0.05
$ & $
3.4\pm 0.9
\cite{PDG}$ &
$ 0 $ & $ --$ \\
\hline
$ \Xi_{c}^{0A}\rightarrow \Xi^{-} \pi^{+} $ & $0.17$ & $1.6 $ & $ --$
 & $--$ &
$ 0.06 $ & $ --$ \\
\hline
$ \Xi_{c}^{+A}\rightarrow \Xi^{0} \pi^{+} $ & $0.88$ & $31$ & $ --$ & $--$ &
$ 0.03 $ & $ --$\\
\hline
$ \Xi_{c}^{0A}\rightarrow \Xi^{0} \pi^{0} $ & $0.62$ & $ 6.1$ & $ --$ & $--$ &
$ -0.89 $ & $ --$\\
\hline
$ \Xi_{c}^{+A}\rightarrow \ \Sigma^{+} {\bar K}^{0} $ & $0.31$ & $
 3.1$ & $ --$ & $--$ &
$ -0.005 $ & $ --$ \\
\hline
$ \Xi_{c}^{0A}\rightarrow \ \Lambda {\bar K}^{0} $ & $0.42$ & $ 4.1$
& $ --$ & $--$ &
$ -0.76 $ & $ --$\\
\hline
$ \Xi_{c}^{0A}\rightarrow \ \Sigma^{0} {\bar K}^{0} $ & $0.23$ & $
 2.2$ & $ --$ & $--$ &
$ 0.006 $ & $ --$\\
\hline
$ \Xi_{c}^{0A}\rightarrow \ \Sigma^{+} { K}^{-} $ & $0.24$ & $ 2.3$ & $ --$
& $--$ &
$ 0 $ & $ --$\\
\hline
\end{tabular}
\end{center}
\end{table}
\begin{table}
\caption{Decay rates ($\times 10^{11}s^{-1}$), branching ratios
($ \times 10^{-3} $) and asymmetry
predictions for Cabbibo favoured
$B_{i} \rightarrow B_f V $ decays  }
\begin{center}
\begin{tabular}{|c|c|c|c|c|c|c|}
\hline
$ Process$ & $\Gamma_{Th}$ & $ BR_{Th} $ & $\Gamma_{Expt}$ & $ BR_{Expt}$ &
$\alpha_{Th}$ & $\alpha_{Expt}$\\
\hline
$ \Lambda_c \rightarrow \Lambda \rho^{+} $ & $0.55$ & $ 11$ &
 $ < 2.1$ & $ < 42 \cite{cleo 2}$ &
$ 0.46 $ & $ --$ \\
\hline
$ \Lambda_c \rightarrow \Sigma^{0} \rho^{+} $ & $0.15$ & $ 3$ & $ --$ & $--
$ & $ 0.0 $ & $ -- $ \\
\hline
$ \Lambda_c \rightarrow \Sigma^{+} \rho^{0} $ & $0.15$ & $ 3$ & $ < 0.6
$ & $ < 12
 \cite{PDG}$ &
$ 0 $ & $ --$ \\
\hline
$ \Lambda_c \rightarrow p { K}^{*0} $ & $0.57$ & $ 11.3$ & $ --$ & $--$ &
$ 0.45 $ & $ --$ \\
\hline
$ \Lambda_c \rightarrow \Xi^{0} { K}^{*+} $ &
 $0.002$ & $0.8 $& $ --$ & $--$ &
$ 0 $ & $ --$ \\
\hline
$ \Xi_{c}^{0A}\rightarrow \Xi^{-} \rho^{+} $ & $1.3$ & $ 12.8$ & $ --$
 & $--$ &
$ 0.54 $ & $ --$ \\
\hline
$ \Xi_{c}^{+A}\rightarrow \Xi^{0} \rho^{+} $ & $0.88$ & $ 31$ & $ --$ &$--$ &
$ 0.46 $ & $ --$\\
\hline
$ \Xi_{c}^{0A}\rightarrow \Xi^{0} \rho^{0} $ & $0.11$ & $ 1.1$ & $ --$ &
 $--$ &
$ 0 $ & $ --$\\
\hline
$ \Xi_{c}^{+A}\rightarrow \ \Sigma^{+} { K}^{*0} $ & $0.36$ & $ 12.8$ & $ --$ &
 $--$ &
$ 0.47 $ & $ --$\\
\hline
$ \Xi_{c}^{0A}\rightarrow \ \Lambda { K}^{*0} $ & $0.10$ & $ 1$ & $ --$ &
 $--$ &
$ -0.56 $ & $ --$\\
\hline
$ \Xi_{c}^{0A}\rightarrow \ \Sigma^{0} { K}^{*0} $ & $0.17$ & $ 1.7$ & $ --$ &
 $--$ &
$ 0.37 $ & $ --$\\
\hline
$ \Xi_{c}^{0A}\rightarrow \ \Sigma^{+} { K}^{*-} $ & $0.016$ & $
0.15$& $--$
 & $ --$ &
$ 0 $ & $ --$\\
\hline
\end{tabular}
\end{center}
\end{table}
\begin{table}
\caption{Decay rate ($\times 10^{11}s^{-1}$), branching ratio relative to
$ p K^{-} \pi^{+} $ mode and asymmetry
predictions for $\Lambda_c\rightarrow p \phi$    }
\begin{center}
\begin{tabular}{|c|c|c|c|c|}
\hline
$ Process$ & $\Gamma_{Th},BR$ & $\Gamma_{Expt}$ & $\alpha_{Th}$ &
$\alpha_{Expt}$\\
\hline
$ \Lambda_c\rightarrow p \phi$  & $0.02,BR \approx 0.01$
 & $-- $ &
$ 0.31$ & $ --$ \\
\hline
\end{tabular}
\end{center}
\end{table}\begin{table}
\caption{Decay rate ($\times 10^{11}s^{-1}$), branching ratio relative to the
total decay width
 and asymmetry predictions for
$\Lambda_b \rightarrow J/\psi  \Lambda $ decays   }
\begin{center}
\begin{tabular}{|c|c|c|c|c|}
\hline
$ Process$ & $\Gamma_{Th}$ & $\Gamma_{Expt}$ & $\alpha_{Th}$ &
$\alpha_{Expt}$\\
\hline
$ \Lambda_b \rightarrow J/\psi \Lambda $ & $0.4\times 10^{-3},BR \approx 0.4
\times 10^{-4}$
 & $-- $ &
$ 0.25$ & $ --$ \\
\hline
\end{tabular}
\end{center}
\end{table}
\begin{table}
\caption{Form factors at the point $q^{2}_{max}$ for $\Lambda_c \rightarrow
\Lambda$    }
\begin{center}
\begin{tabular}{|c|c|c|c|c|c|}
\hline
$f_1(q^{2}_{max})$ & $f_2(q^{2}_{max})$ & $f_3(q^{2}_{max})$ &
$g_1(q^{2}_{max})$ & $g_2(q^{2}_{max})$ & $g_3(q^{2}_{max})$\\
\hline
$0.46 $ & $0.11$ &$-0.84 $ & $0.46$ & $0.26 $ & $-1.88$ \\
\hline
\end{tabular}
\end{center}
\end{table}

\begin{table}
\caption{Form factors at the point $q^{2}=0$ for $\Lambda_c \rightarrow
\Lambda$    }
\begin{center}
\begin{tabular}{|c|c|c|c|c|c|}
\hline
$f_1(q^{2}=0)$ & $f_2(q^{2}=0)$ & $f_3(q^{2}=0)$ &
$g_1(q^{2}=0)$ & $g_2(q^{2}=0)$ & $g_3(q^{2}=0)$\\
\hline
$0.22 $ & $0.05$ &$-0.40 $ & $0.28$ & $0.16 $ & $-1.15$ \\
\hline
\end{tabular}
\end{center}
\end{table}
\begin{table}
\caption{Form factors at the point $q^{2}_{max}$ for $\Lambda_b \rightarrow
\Lambda$    }
\begin{center}
\begin{tabular}{|c|c|c|c|c|c|}
\hline
$f_1(q^{2}_{max})$ & $f_2(q^{2}_{max})$ & $f_3(q^{2}_{max})$ &
$g_1(q^{2}_{max})$ & $g_2(q^{2}_{max})$ & $g_3(q^{2}_{max})$\\
\hline
$0.38 $ & $0.11$ &$-0.29 $ & $0.46$ & $0.22 $ & $-0.0175 $ \\
\hline
\end{tabular}
\end{center}
\end{table}
\begin{table}
\caption{Form factors at the point $q^{2}=0$ for $\Lambda_b \rightarrow
\Lambda$    }
\begin{center}
\begin{tabular}{|c|c|c|c|c|c|}
\hline
$f_1(q^{2}=0)$ & $f_2(q^{2}=0)$ & $f_3(q^{2}=0)$ &
$g_1(q^{2}=0)$ & $g_2(q^{2}=0)$ & $g_3(q^{2}=0)$\\
\hline
$0.040 $ & $0.012$ &$-0.03 $ & $0.08$ & $0.04 $ & $-0.003$ \\
\hline
\end{tabular}
\end{center}
\end{table}
\newpage
In conclusion we have studied the non-leptonic two body decays of
charmed and bottom baryons involving transition of a heavy to light quark
based on a model for form factors that includes $1/m_Q$ corrections.

{\bf Acknowledgement } :
I would like to thank Professor Sandip Pakvasa and Professor Tom Browder
 for useful discussions. This
 work was supported in part by US
D.O.E grant \# DE-FG 03-94ER40833.

\end{document}